\documentclass[a4paper,fleqn]{cas-dc}

\usepackage[authoryear]{natbib} % Use natbib with author-year format
\usepackage{xcolor}             % For text coloring


\usepackage[utf8]{inputenc}      % Allows UTF-8 input
\usepackage{multirow}
\usepackage{longtable}
\usepackage{booktabs}
\usepackage{tabularx}
\usepackage{graphicx}
\usepackage{amssymb,amsmath}
\usepackage{hyperref}
\let\oldAA\AA
\renewcommand{\AA}{\text{\normalfont\oldAA}}
\usepackage{pdflscape}

\def\tsc#1{\csdef{#1}{\textsc{\lowercase{#1}}\xspace}}
\tsc{WGM}
\tsc{QE}
\newcommand{\Msun}{\ensuremath{M_\odot}}         % Solar mass
\newcommand{\lambdaEdd}{\ensuremath{\lambda_{\mathrm{Edd}}}} % Eddington ratio
\def\aj{AJ}%
\def\araa{ARA\&A}%
\def\apj{ApJ}%
\def\apjl{ApJ}%
\def\apjs{ApJS}%
\def\aap{A\&A}%
\def\aapr{A\&A~Rev.}%
\def\mnras{MNRAS}%
\def\nat{Nature}%
\def\prd{Phys.~Rev.~D}%
\def\prl{Phys.~Rev.~Lett.}%
\def\pasp{PASP}%
\def\nar{New~Astron.~Rev.}%
\begin{document}

%%%%%%%%%%%%%%%%%%%%%%%%%%%%%%%%%%%%%%%%%%%%%%%
\let\WriteBookmarks\relax
\def\floatpagepagefraction{1}
\def\textpagefraction{.001}

% Short title
\shorttitle{Quasars Growth}    

% Short author
\shortauthors{Yu Wang \& Remo Ruffini}

% Main title of the paper
%\title [mode = title]{Multi-wavelength temporal and spectral study of the FSRQ B2\,1348+30B}  
\title [mode = title]{Growth of High-Redshift Quasars from Fermion Dark Matter Seeds}  

% Title footnote mark
% eg: \tnotemark[1]
\tnotemark[<tnote number>] 

% Title footnote 1.
% eg: \tnotetext[1]{Title footnote text}
%\tnotetext[<tnote number>]{<tnote text>} 

% First author
%
% Options: Use if required
% eg: \author[1,3]{Author Name}[type=editor,
%       style=chinese,
%       auid=000,
%       bioid=1,
%       prefix=Sir,
%       orcid=0000-0000-0000-0000,
%       facebook=<facebook id>,
%       twitter=<twitter id>,
%       linkedin=<linkedin id>,
%       gplus=<gplus id>]

\author[1,2,3]{Yu Wang}

% Corresponding author indication
%\cormark[1]

% Footnote of the first author
%\fnmark[<footnote mark no>]

% Email id of the first author
\ead{yu.wang@icranet.org}

% URL of the first author
%\ead[url]{<URL>}

% Credit authorship
% eg: \credit{Conceptualization of this study, Methodology, Software}
%\credit{<Credit authorship details>}

% Address/affiliation
\affiliation[1]{organization={ICRA, Dip. di Fisica, Sapienza Universit\`a  di Roma},
            addressline={Piazzale Aldo Moro 5}, 
            city={Roma},
%          citysep={}, % Uncomment if no comma needed between city and postcode
            postcode={I-00185}, 
       %    state={},
            country={Italy}}
            
\affiliation[2]{organization={ICRANet},
            addressline={Piazza della Repubblica 10}, 
            city={Pescara},
%          citysep={}, % Uncomment if no comma needed between city and postcode
            postcode={65122}, 
       %    state={},
            country={Italy}}
            
 \affiliation[3]{organization={INAF -- Osservatorio Astronomico d'Abruzzo},
            addressline={Via M. Maggini snc}, 
            city={Teramo},
%          citysep={}, % Uncomment if no comma needed between city and postcode
            postcode={I-64100}, 
       %    state={},
            country={Italy}}

\author[1,2,4,5]{Remo Ruffini}

% Footnote of the second author
%\fnmark[]

% Email id of the second author2009ApJ...692...32D
\ead{ruffini@icra.it}

\affiliation[4]{organization={Universit\'e de Nice Sophia-Antipolis},
            addressline={Grand Ch\^ateau Parc Valrose}, 
            city={Nice},
%          citysep={}, % Uncomment if no comma needed between city and postcode
            postcode={CEDEX 2}, 
          %  state={Maharashtra},
            country={France}}
            
\affiliation[5]{organization={INAF},
            addressline={Viale del Parco Mellini 84}, 
            city={Roma},
%          citysep={}, % Uncomment if no comma needed between city and postcode
            postcode={00136}, 
            %state={Maharashtra},
            country={Italy}}   
            
%\cormark[1]     
% URL of the second author
%\ead[url]{}

% Corresponding author indication

% Credit authorship
%\credit{}

% Corresponding author indication
%\cormark[5]

% Footnote of the first author
%\fnmark[<footnote mark no>]

% URL of the first author
%\ead[url]{<URL>}
%\credit{}

\cortext[1]{Corresponding author}
% Credit authorship
% eg: \credit{Conceptualization of this study, Methodology, Software}
%\credit{<Credit authorship details>}

% Address/affiliation

% Footnote text
%\fntext[1]{}

% Abstract of the paper
\begin{abstract}
Quasars hosting $\gtrsim 10^{9}\,M_\odot$ black holes at $z>6$ challenge growth scenarios that start from light seeds and assume accretion within already formed galaxies. Motivated by the James Webb Space Telescope (JWST) discovery of Little Red Dots (LRDs), which suggests that $\sim 10^{6}\,M_\odot$ black holes can be active in compact, pre-galactic environments, we revisit early black hole growth with a minimal cosmology-based framework. We model the accretion history as the smaller of the Bondi inflow rate and the Eddington-limited rate, where the Bondi rate is set by the supply of overdense primordial gas whose density declines with cosmic expansion, and the Eddington rate captures regulation by radiative feedback. By fitting the observed masses and luminosities of J0313--1806 ($z=7.64$) and J0100+2802 ($z=6.30$) with Bayesian inference, we infer initial conditions that favor massive seed black holes with initial mass $M_0 \sim 10^{6}\,M_\odot$, formed at $z\sim20{-}30$ in environments with baryonic overdensity factors $f_\rho \gtrsim 50$ relative to the cosmic mean. The resulting growth histories include a prolonged supply-limited stage, and they reproduce the observed quasar masses without requiring sustained Eddington accretion or any super-Eddington episodes. The inferred seed mass scale is consistent with black holes produced by the collapse of quantum-degenerate fermion dark matter cores, providing a physically defined pathway to massive seeds at the redshifts implied by LRD phenomenology.
\end{abstract}


\maketitle

%%%%%%%%%%%%%%%%% BODY OF PAPER %%%%%%%%%%%%%%%%%%
\section{Introduction} \label{sec:intro}
The discovery of 3C 273 by Martin Schmidt in 1963 first established that quasars are cosmological sources, with point-like images corresponding to extreme distances rather than compact physical sizes \citep{1963Natur.197.1040S}. The measured redshift, $z = 0.158$, implied a bolometric luminosity $L_{\text{bol}} \gtrsim 10^{46}~\text{erg}~\text{s}^{-1}$, confirming that quasars are among the most energetic objects in the universe. Although the cosmological origin of quasars was clear from the start, their physical nature remained uncertain.

Early models sought to explain the extreme luminosities by invoking accretion onto massive compact objects. In 1964, Salpeter proposed that accretion onto a $10^6M_\odot$ core with general relativistic properties as discussed by Feynman could power a quasar \citep{1964ApJ...140..796S}. Around the same time, Zel'dovich discussed the role of gravitational collapse of a $10^6$--$10^7M_\odot$ core, following the Oppenheimer--Snyder model, as a mechanism for quasar activity \citep{1964SPhD....9..195Z}.

The first Texas Symposium on Relativistic Astrophysics in 1963 \citep{1989PhT....42h..46S} presented early quasar observations, prompting new proposals for their physical origin. A critical theoretical advance was the discovery in 1963 of the exact solution to Einstein's equations for a rotating star by Roy Kerr \citep{1963PhRvL..11..237K}.  By 1967, the possibility of extracting rotational energy from a Kerr metric for astrophysical phenomena was proposed by Ruffini and Wheeler, who formulated the general equation for energy extraction in the ``ergosphere'' \citep{1971ESRSP..52...45R}. These developments converged in the presentation of the concept of the ``black hole'', introduced in Physics Today \citep{1971PhT....24a..30R}. That black hole in Cygnus X-1 was first proved by Ruffini at the Texas Symposium in New York \citep{hegyi1974sixth}. Today, black hole formation is routinely observed in association with gamma-ray bursts (GRBs) of $> 2.35 M_\odot$ \citep{2021MNRAS.504.5301R}.

Historically, black holes were primarily regarded as endpoints of massive star evolution. This made it difficult to imagine a role for black holes in the early universe, where the physics appeared to be dominated by relativistically expanding processes. Nevertheless, accumulating evidence from cosmological observations and theory has motivated a shift in perspective.

The interest toward the earliest epoch of the universe began in Los Alamos in 1945, through an exceptional collaboration between Gamow \citep{1948PhRv...74..505G, 1948Natur.162..680G},  Fermi and Turkevich \citep[mentioned in][]{1950RvMP...22..153A, 1953ARNPS...2....1A}. They proposed that the early universe was radiation-dominated and established a quantitative framework linking light-element abundances to the cosmic expansion rate. These works provided the basis for the theory of primordial nucleosynthesis: at redshift $z \sim 10^9$, the photon temperature was $T \approx 10^9$ K and the baryon density was $\rho_b \sim 10^5~\text{g}~\text{cm}^{-3}$, only light nuclei up to $^7$Li could be produced. By 1965, these theoretical works had been almost forgotten, with the sole exception of a few scientists \citep[details in][]{ruffini_inpress_book}. They were dramatically revived by the discovery of the cosmic microwave background (CMB) radiation by Penzias and Wilson \citep{1965ApJ...142..419P}. See Fig. \ref{fig:z-evolution} for the density evolution of baryons and radiation at different redshifts.

A major breakthrough in galactic dynamics came from the study of flat rotation curves in spiral galaxies. Starting in the late 1960s \citep{1978ApJ...225L.107R} and extending through the 1980s \citep{1981AJ.....86.1791B, 1981AJ.....86.1825B, 1985ApJ...295..305V, 1987IAUS..117...67S}, the work of Vera Rubin and others demonstrated that observed rotation curves could not be explained by baryonic matter alone. This indicated the need for a substantial dark matter component, potentially linked to the distribution of baryonic and non-baryonic mass.

The search for a Kerr black hole by ESO in the Galactic Center was initiated by Townes and Genzel in 1977, employing mid-infrared spectroscopy of ionized gas near Sgr A$^*$ at La Silla, which revealed velocity dispersions consistent with a $\sim 3\times10^{6}\ M_\odot$ compact object \citep{1987ARA&A..25..377G}, and has continued for more than 50 years \citep{1994RPPh...57..417G, 2010RvMP...82.3121G, 2018A&A...618L..10G,2018A&A...615L..15G,2024A&ARv..32....3G}.

These developments matured the RAR model, in which the equilibrium configurations of self-gravitating fermions are described by solutions to the Einstein--Fermi--Dirac system \citep{2015MNRAS.451..622R}. The most general solutions predict three main regions in DM-dominated galaxies: (i) an inner quantum-degenerate core of nearly constant density, (ii) an intermediate region where the DM density decreases sharply, and (iii) an outer, classical Boltzmann-like regime overlapping with baryonic components. The first astrophysical evidence for such structure came from observations at the Galactic Center, particularly through the long-term study of S-star orbits around Sgr A$^*$. Although the search for a Kerr black hole in Sgr A$^*$ remains inconclusive \citep{2018A&A...618L..10G,2018A&A...615L..15G}, the relativistic motion of S-stars is consistent with the presence of a self-gravitating dark matter core as predicted by the RAR model for fermion masses in the range 56--378 keV \citep{2020A&A...641A..34B, 2021MNRAS.505L..64B, 2021MNRAS.502.4227A}.

Following these results, a model was proposed in which a neutral fermion dark matter component (named ``X-fermion'') is responsible for structure formation, leading to the formation of supermassive cores in galaxy centers \citep{ruffini2025role}. The critical mass for gravitational collapse into a black hole for a degenerate fermion gas is given by
\begin{equation}
M_{\text{cr}} \simeq 6.95 \times 10^6 \left( \frac{300~\mathrm{keV}}{m_X} \right)^2 M_\odot,
\end{equation}
where $m_X$ is the fermion mass. Requiring that the compact object in the Galactic Center has not yet collapsed into a black hole imposes an upper bound on $m_X < 381$ keV, and sets a lower limit for the mass of any cosmological DM supermassive black hole (SMBH) as
\begin{equation}
M_{\text{SMBH}} > M_{\mathrm{SgrA}^*} = 4.3 \times 10^6~M_\odot.
\label{eq:1e6}
\end{equation}

In this article, motivated by the recent discovery of a population of compact, red sources known as LRDs \citep{2024ApJ...963..129M,2025arXiv250917484R}, we consider a seed black hole mass for quasar formation similar to that proposed by Salpeter \citep{1964ApJ...140..796S}, but with a crucial difference: rather than accretion of baryonic matter onto a black hole moving within a pre-existing galaxy, our scenario focuses on the accretion of primordial cosmological matter onto black hole seeds formed by the collapse of dark matter cores in a pre-galactic environment.

Explaining the rapid formation of billion solar masses black holes by $z \sim 7$ often involves theoretical models assuming continuous accretion at or near the Eddington limit to maximize growth rates \citep{2020ApJ...897L..14Y,2020ARA&A..58...27I,2021NatRP...3..732V}. However, such assumptions may oversimplify the evolving cosmological environment, in particular the rapid dilution of the ambient gas density driven by cosmic expansion, which is one of the main points of this article. In addition, LRDs provide observational motivation to revisit the earliest growth stage of AGN, because they suggest that massive black hole seeds can already be formed while the surrounding system is still dominated by primordial gas, before a mature galactic potential, disk, or bulge has formed. In this work we treat LRDs as the empirical starting point, and we explore the possibility that dark matter black hole seeds, formed as early as redshift $z \sim 20-30$, were exposed to rapidly accreting primordial gas, originating from the Big Bang, before the formation of galaxies \citep{2025ApJ...983...60C}. This constitutes the second major point of this article.

Based on the above two points, this work adopts a more physically grounded yet simple approach. We place a black hole seed within a region of primordial gas, and we use a deliberately simple, cosmology based growth model to estimate AGN mass build-up during the LRD-like pre-galactic phase, rather than modeling accretion within an already formed galaxy structure. We explore how an overdense primordial region can supply gas to the seed, enabling rapid growth toward quasar masses, as explained in Section \ref{sec:basic-picture}.

In the following Section \ref{sec:agn-growth}, we present the black hole growth framework by modeling accretion as the minimum of the Eddington and Bondi rates. In Section \ref{subsec:accretion}, we describe the physical derivation of these accretion prescriptions and their cosmological evolution. In Section \ref{subsec:quasar-samples}, we introduce the observed quasar samples used as constraints. In Section \ref{subsec:fitting-methods}, we detail our Bayesian Markov Chain Monte Carlo fitting method to infer the initial seed mass and overdensity. In Section \ref{subsec:results}, we analyze the evolutionary phases, initial Eddington-limited growth, a transitional Bondi phase, and a late resurgence of Eddington growth, as determined by the competition between black hole mass increase and ambient baryon dilution. Section \ref{sec:discussion} provides a comparison between our scenario and alternative seeding models such as Population III remnants and direct collapse black holes. In Section \ref{sec:conclusion}, we summarize our findings and describe how the present results will be extended in follow-up studies based on cosmological perturbation theory and a unified formation picture for quasars and LRD.

\section{Basic Picture}\label{sec:basic-picture}

The discovery of early galaxies and massive black holes by the James Webb Space Telescope (JWST) has fundamentally changed our understanding of the early universe. Notably, JWST discovered LRDs, which constitute more than $10\%$ of observed galaxies at high redshift but were not predicted by previous cosmological simulations \citep{2024ApJ...968...34W,2025ApJ...978...92L,2025ApJ...986..126K,2025ApJ...979..138H}.

Several hundred LRDs have been observed, existing within the first few hundred million years after the Big Bang. These objects have sizes $\sim 10\%$ that of the Milky Way and may harbor black holes with masses on the order of millions of solar masses \citep{2024RNAAS...8..207G}. \citet{2025ApJ...983...60C} attempted to identify host galaxies for the LRDs but found only one associated with an LRD at redshift $z = 4.96$, with a size less than 1 kpc. No host galaxies were found for the other seven LRDs. Moreover, the mass of the central black holes in LRDs accounts for $5\%$ to $40\%$ of the stellar mass, far above the $\sim 0.1\%$ ratio that characterizes local galaxies \citep{2025ApJ...983...60C,2025arXiv250613852M}.

LRDs may be candidates for the first generation of galaxies or proto-galaxies \citep{2025ApJ...984..121W,2025A&A...698A.317M}. These observations have several immediate consequences. First, the central black hole may form before the galaxy itself. Second, the structure of LRDs appears to be much simpler than that of mature galaxies, with minimal feedback from stellar evolution on the galaxy structure. Third, both the central black hole and the host galaxy grow in the early universe. In this sense, LRDs can be considered as ``naked black holes'' accreting surrounding gas.

The historical developments introduced in Sec. \ref{sec:intro} and current JWST observations motivate us to focus on key discoveries and to propose simple hypotheses, without resorting to detailed modeling or computation, in order to reconsider the problem of the formation and growth of massive black holes. Therefore, we simplify the model as follows: the black hole seed accretes surrounding gas in an overdense region before the formation of galaxy.

There are several proposed mechanisms for the formation of black hole seeds, including the collapse of Population III stars and direct collapse of baryonic matter (DCBH) \citep{2010A&ARv..18..279V}. The simple model presented here does not exclude any specific channel for seed formation. However, based on previous studies, such as UHZ-1 \citep{2024NatAs...8..126B}, the supermassive black holes with masses $\sim 10^8 M_\odot$ at high redshift requires seed masses of at least $10^4$–$10^5 M_\odot$ and keeps accreting at Eddington-limited rate. If sustained Eddington accretion cannot be maintained, the required initial seed mass must be even higher, often above $10^5 M_\odot$. Such high seed masses are difficult to achieve through standard baryonic mechanisms alone. Collapse of dark matter, therefore, provides an additional and potentially viable pathway \citep{ruffini2025role, 2025arXiv250323710J}. The properties of dark matter, both its particle type and energy range, remain largely unconstrained. In this work, we consider $\sim 300$ keV fermion dark matter, which has been shown to fit both the Galactic rotation curve and the orbits of the S2 cluster around Sgr $A^*$ \citep{2020A&A...641A..34B, 2021MNRAS.505L..64B, 2021MNRAS.502.4227A}. The critical mass for such dark matter is of order $10^6 M_\odot$, consistent with the requirement for forming massive seeds \citep{ruffini2025role}. Nonetheless, we do not rule out the possibility of other dark matter candidates that can satisfy these constraints.

Based on the basic picture, the model has two key parameters: the initial mass of a black hole seed and the density of the surrounding baryonic matter. Observational constraints are provided by the measured masses and luminosities of high-redshift quasars. To quantitatively match these observations, it is also necessary to specify the onset time of accretion onto the black hole. In this study, we do not impose any a priori constraint on the accretion start time. In this study, we do not impose any a priori constraint on the accretion start time. Instead, we calculate the black hole growth history for a range of initial redshifts and determine the redshift at which an initial mass of $\sim 10^6~M_\odot$, set by $\sim 300$ keV fermions, reproduces the observed properties of high-redshift quasars.

%One detail to note is that our calculation only considers the accretion of baryonic matter. The black hole seed may undergo an initial, short phase of dark matter-dominated accretion at formation. However, in this work, we define the onset time as the beginning of baryonic matter-dominated accretion, which is the stage that governs the subsequent rapid growth of the black hole and determines its observable properties.

\section{AGN Growth}\label{sec:agn-growth}

We model the growth governed by the minimum of the Eddington limit and the density dependent Bondi rate, the dominance of Eddington or Bondi accretion depends on whether radiation pressure (Eddington limit) or gas supply (Bondi rate) imposes a tighter constraint, accounting for black hole mass and cosmic expansion at each epoch. 

By applying this framework to fit the observed masses and luminosities of well-characterized high-redshift quasars, such as J0313-1806 \citep{2021ApJ...907L...1W} and J0100+2802 \citep{2015Natur.518..512W} detailed below, we can trace their evolutionary histories backward in time. This methodology, employing Bayesian inference, allows us to constrain the plausible initial conditions, specifically the required seed mass ($M_0$) and the environmental density factor ($f_\rho$), necessary to produce these SMBHs by their observed epochs. As we will demonstrate, the results consistently favor massive seeds ($M_0 \sim 10^6 \, M_\odot$) originating in overdense regions ($f_\rho \gtrsim 50-100$), lending  support to seeding scenarios, as the gravitational collapse of dense fermion cores, which naturally predict such initial conditions.

\subsection{Accretion}\label{subsec:accretion}

%Once the degenerate X-fermion core has collapsed into a $\sim10^{6}$ black hole, the new object sits in the center of a halo of mass $> 10^{8} M_\odot$. That halo contains a baryonic reservoir $> 10^{7} M_\odot$ \citep{2024ApJ...961L..10A}. To estimate how rapidly this gas can feed the black hole we compare the Bondi inflow rate, set by local gas properties, with the Eddington-limited rate, set by radiation feedback.

Bondi accretion describes the process by which a massive object, such as a black hole, accretes matter from a surrounding gas \citep{1952MNRAS.112..195B}. For a stationary black hole in a uniform gas, the Bondi accretion rate, $\dot{M}_B$, is given by

\begin{equation}
\dot{M}_B = 4\pi \frac{(G M_{\text{BH}})^2 \rho}{c_s^3},
\end{equation}
where $G$ is the gravitational constant, $M_{\text{BH}}$ is the black hole mass, and $c_s$ is the sound speed of the gas. The local gas density is parameterized as $\rho = f_\rho \rho_{b}$, where $\rho_{b}$ is the average baryon density of the universe and $f_\rho \gg 1$ denotes the local overdensity. Because the processes considered here occur in low-mass halos at early times, the virial velocity is lower than in massive galaxies. As a result, the sound speed $c_s$ (or the relative inflow velocity) is also expected to be low. We adopt a typical fiducial value of $10~\text{km s}^{-1}$ \citep{2004NewAR..48..843E}. While this velocity may change at different time, it is expected to remain within the same order of magnitude during the period of interest. For simplicity, we keep $c_s$ constant in our calculations.

The Eddington luminosity imposes a limit on the accretion rate of a black hole when the radiation pressure from the accreting material balances gravitational infall. This occurs because the photon flux ionizes surrounding gas, creating electron scattering opacity, which drives winds that counteract accretion \citep{1979rpa..book.....R}. The Eddington accretion rate is

\begin{equation}
\dot{M}_{\text{Edd}} = \frac{L_{\text{Edd}}}{\eta c^2} = \frac{4\pi G M_{\text{BH}} m_p}{\eta \sigma_T c},
\end{equation}
where $L_{\text{Edd}}  = 4\pi G M_{\text{BH}} c m_p / \sigma_T$ is the Eddington luminosity, $m_p$ is the proton mass, $\sigma_T$ is the Thomson scattering cross-section, $c$ is the speed of light, and $\eta$ is the accretion efficiency, normally taken as $0.1$ \citep{2003ApJ...582..133D}. In high-density environments, radiation feedback regulates accretion to $\dot{M}_{\text{Edd}}$, as exceeding this would unbind the gas. 

\begin{figure}
    \centering
    \includegraphics[width=1\linewidth]{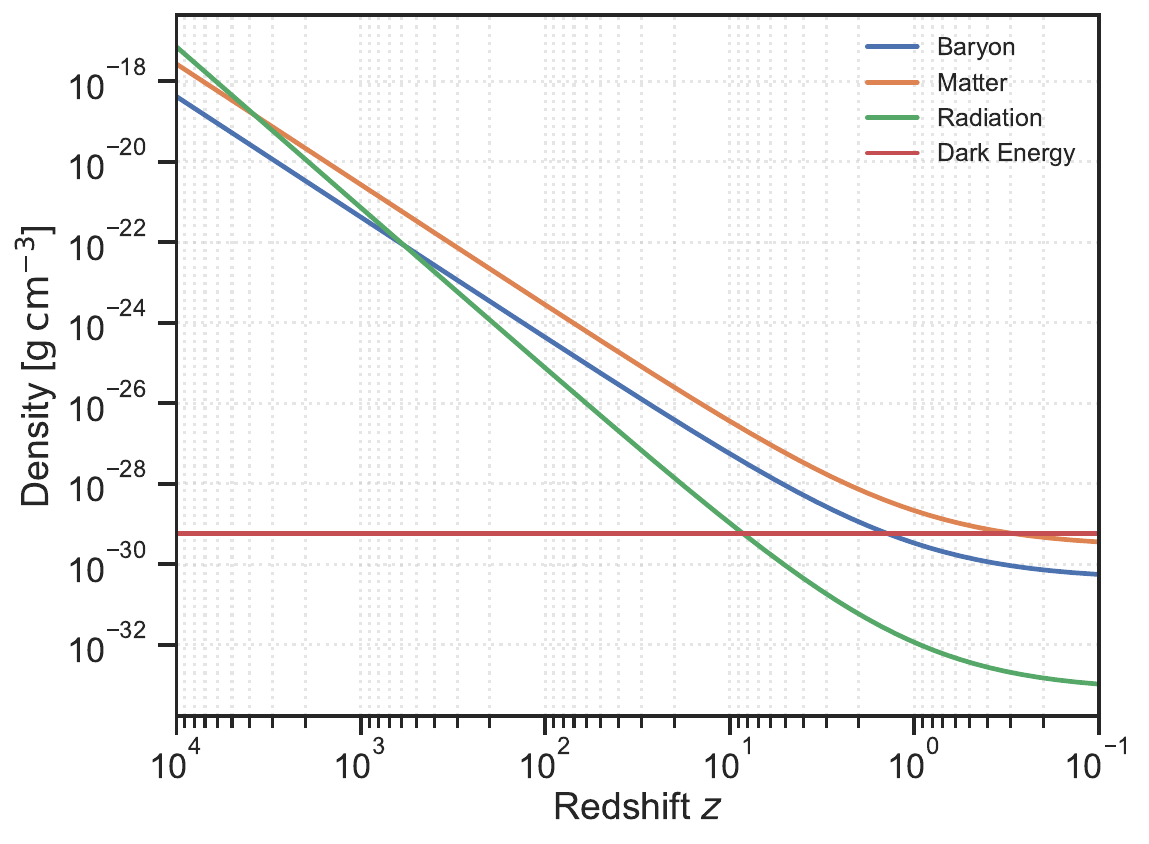}
    \caption{Redshift evolution of the physical mass–energy density of the four standard cosmological components, computed with the Planck 2018 parameters $H_{0}=67.4\;\text{km\,s}^{-1}\text{Mpc}^{-1}$, $\Omega_{\mathrm b}=0.049$, $\Omega_{\mathrm m}=0.315$, $\Omega_{r}=8.4\times10^{-5}$ and $\Omega_{\Lambda}=0.685$.  The present-day critical density is $\rho_{\mathrm{crit},0}=8.5\times10^{-30}\,\text{g\,cm}^{-3}$, giving $\rho_{\mathrm b,0}=4.2\times10^{-31}\,\text{g\,cm}^{-3}$, $\rho_{\mathrm m,0}=2.7\times10^{-30}\,\text{g\,cm}^{-3}$, $\rho_{r,0}=7.2\times10^{-34}\,\text{g\,cm}^{-3}$ and $\rho_{\Lambda,0}=5.8\times10^{-30}\,\text{g\,cm}^{-3}$.  Densities are followed from $z=10^{4}$ to $z=0.1$ assuming the scalings $\rho\propto(1+z)^{3}$ for baryons and total matter, $\rho\propto(1+z)^{4}$ for radiation, and $\rho=$ constant for dark energy.
}
    \label{fig:z-evolution}
\end{figure}

\begin{figure}
    \centering
    \includegraphics[width=1\linewidth]{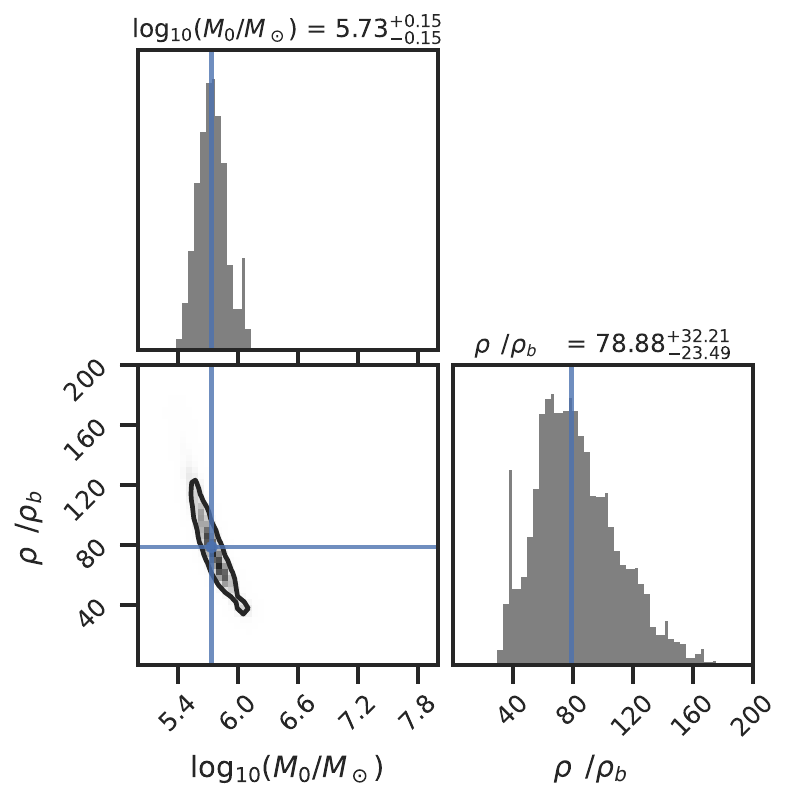}
\caption{
\textbf{Example showing the Monte Carlo inference of initial conditions for quasar J010013.02+280225.8. growth.}
The posterior probability distributions of the two key parameters in our model: initial black hole mass (log${10}(M_0/M_\odot) = 5.73^{+0.15}_{-0.15}$) at $z = 1100$ and ambient density enhancement ($\rho/\rho_{\text{b}} = 78.88^{+32.21}_{-23.49}$). Upper panels show marginalized posterior distributions for each parameter. The lower left panel displays the joint probability distribution with 1-$\sigma$ contours, revealing the degeneracy between initial mass and density, a more massive seed can grow in a less dense environment, while a lighter seed requires higher density to reach the observed final mass. The right panel shows the marginalized posterior for the density enhancement, indicating that significant overdensities ($>50$ times cosmic average) are required for efficient early growth.
}
\label{fig:corner}
\end{figure}

The present-day baryon matter density of the universe is given by $ \rho_{b,0} = \Omega_b \rho_{\text{crit},0} $, where $ \Omega_b = 0.049 $ is the baryon matter density parameter, and $ \rho_{\text{crit},0} = 3H_0^2 / (8\pi G) $ is the critical density, from the Planck 2018 results \citep{2020A&A...641A...6P}. Using a Hubble constant of $ H_0 = (67.4 \pm 0.5)\, \mathrm{km\,s^{-1}\,Mpc^{-1}} $, the critical density evaluates to $ \rho_{\text{crit},0} \simeq 8.5 \times 10^{-30}\, \mathrm{g\,cm^{-3}} $, resulting in a present-day matter density of $\rho_{b,0} = 4.18 \times 10^{-31}\, \mathrm{g\,cm^{-3}}$. From the standard $\Lambda$CDM model, shown in Fig. \ref{fig:z-evolution}, the physical matter density evolves with the scale factor as 
\begin{equation}
\rho_b(z) = \rho_{b,0} (1+z)^3. 
\end{equation}
Therefore, in the very early universe, at redshifts $z \gg 10$, the average matter density was several orders of magnitude higher than at present. The dense gas inflow, which supplies most of the baryonic mass in forming galaxies, would have produced very large Bondi accretion rates, potentially exceeding the Eddington limit by a significant factor. As the universe expanded over billions of years, the average density of matter decreased significantly, consequently, the Bondi accretion rate also decreased and fell below the Eddington limit for not so massive SMBHs. At this point, the limitation imposed by radiation pressure became less stringent, and the actual accretion rate onto the black hole could then be well-approximated by the Bondi accretion rate.

Several simplifying assumptions underlie this accretion framework. First, we neglect the angular momentum of the infalling material and approximate the inflow as spherically symmetric. Our main observational motivation comes from the LRD morphology \citep{2025arXiv250710659L}, where the dominant component is an extended envelope (nicknamed ``The Egg'') rather than an ordered disk. An envelope-dominated structure does not single out a preferred rotation axis, which suggests that the net angular momentum of the gas feeding the central object is small, at least in the time-averaged sense relevant for our growth estimate. A separate line of evidence comes from theory: cosmological simulations typically find only modest halo spin parameters, $\lambda\sim0.03{-}0.05$ \citep{2008MNRAS.388..863C}, and turbulent, multi-stream inflows can further reduce the net angular momentum of the material that actually collapses. Together with the expectation that direct-collapse seeds do not rapidly spin up at these early times, these arguments support a minimal model in which angular momentum and disk formation are sub-dominant, so a Bondi-like, quasi-spherical accretion prescription is an appropriate first-step description of the early growth phase.

Second, we examine whether a power-law gas profile would invalidate the Bondi approximation by enclosing, within the Bondi radius $r_B$, a gas mass larger than the black hole mass. For a spherically averaged density profile $\rho(r)=\rho_b\,(r/r_B)^{-\alpha}$ with $\alpha=1.5$, a seed with $M_{\mathrm{BH}}=5\times10^6\,M_\odot$ at $z=20$ and an overdensity factor $f_\rho=100$ gives an enclosed gas mass $M_{\mathrm{gas}}\simeq4.8\times10^5\,M_\odot$, that is, $M_{\mathrm{gas}}/M_{\mathrm{BH}}\simeq0.095$. Thus, during the early high-redshift phase relevant for seed growth, the black hole gravitational potential dominates, and the Bondi prescription remains valid.

Third, galaxy mergers can transiently increase the gas supply and trigger enhanced accretion. Empirically, merger rates imply that major mergers occur, on average, once per a few hundred Myr to a few Gyr, with the exact rate depending on redshift, galaxy mass, and environment \citep{2025MNRAS.540..774D}. In our model, the net effect of mergers can be encapsulated through the overdensity parameter $f_\rho$. A merger episode can temporarily raise $f_\rho$ and thus the Bondi inflow, shortening the supply-limited phase, but it does not remove the need for a massive seed and a dense environment, it mainly shifts the system more rapidly toward the Eddington-limited regime. The cosmic time elapsed from $z=20$ to $z=10$ is $\simeq 300$~Myr, so within the time window most relevant to our early growth calculations, a merger is possible but not guaranteed. So for simplicity, we ignore the merger effect.

Finally, we do not incorporate a duty cycle into the growth history. The concept of a duty cycle is appropriate for intermittent accretion in mature galaxies, but current LRD samples observationally show no clear evidence for strong jets or large amplitude flaring that would indicate highly intermittent, on-off accretion \citep{2025arXiv250917484R}. In the proto-galactic regime considered here, the accretion rate is primarily controlled by the evolving gas supply set by the ambient density. As we show below by explicitly comparing $\dot{M}_B$ and $\dot{M}_{\mathrm{Edd}}$, the growth is supply-limited for most of the interval from $z\simeq20$ to $z\simeq7$, that is, $\dot{M}_B<\dot{M}_{\mathrm{Edd}}$ for most of the evolution that we are interested in this article. In this regime the radiative output is sub-Eddington and feedback is not expected to produce strong intermittency, while the Eddington-limited phase, when it occurs, occupies only a small fraction of the evlotion for high redsfhit quasars. A constant duty-cycle factor would therefore not capture the time-dependent transition between supply-limited and Eddington-limited growth in our model.

\subsection{Quasar Samples}\label{subsec:quasar-samples}

\begin{figure*}
    \centering
    \includegraphics[width=1\linewidth]{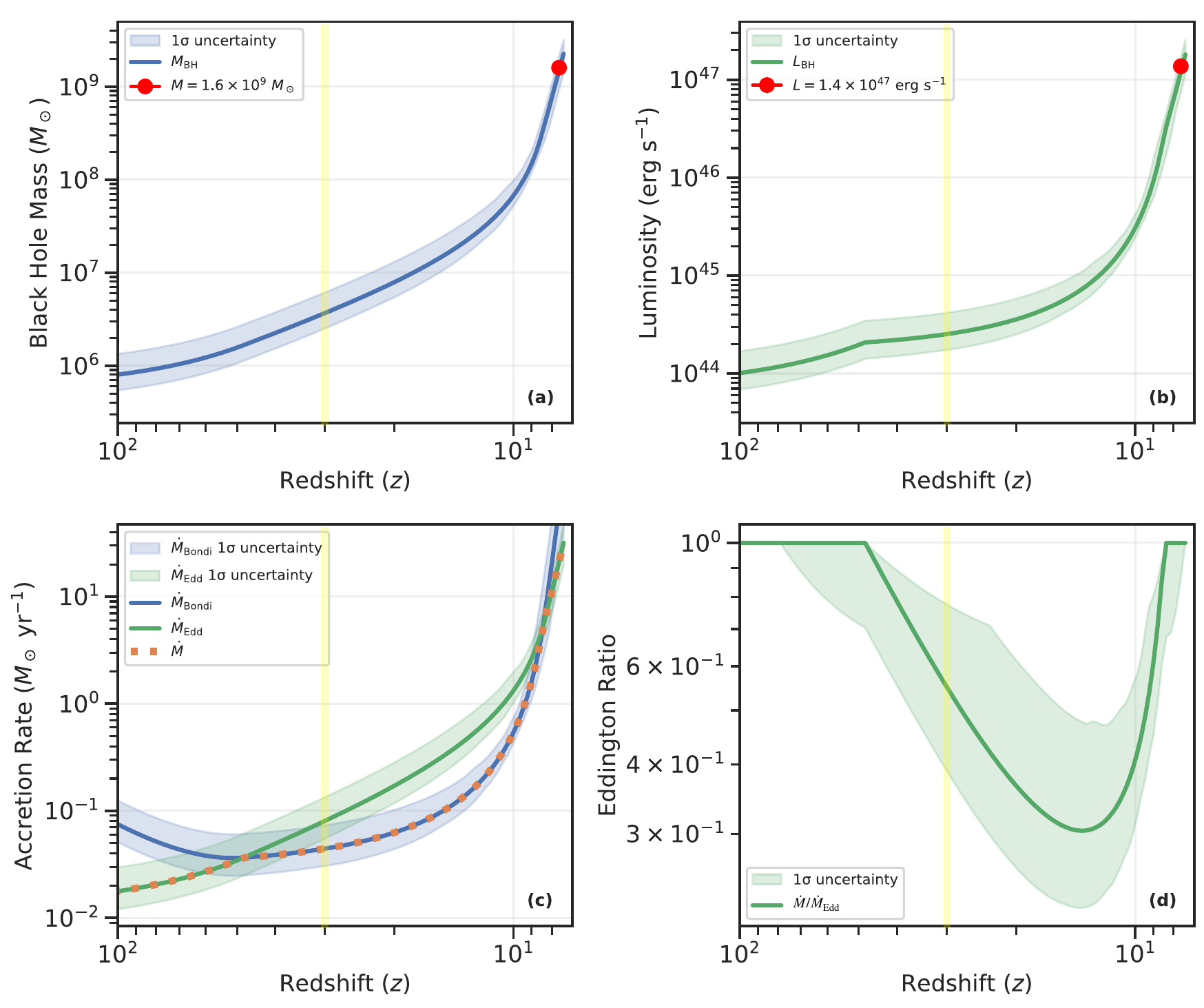}
    \caption{
    \textbf{Evolutionary history of the quasar in J0313-1806.} 
    (a) Black hole mass growth, starting from $9.1^{+4.3}_{-3.6} \times 10^5\, M_\odot$  at  $z = 100$ to  $1.6 \times 10^9\, M_\odot$ at $z = 7.64$ in a dense environment with $\rho_0 / \rho_{\mathrm{b}} = 77.12^{+47.59}_{-29.99}$. 
    (b) Corresponding luminosity evolution, reaching $1.4 \times 10^{47}$ erg s$^{-1}$. 
    (c) Accretion rate history showing the transition between Bondi-limited (blue) and Eddington-limited (green) regimes, with the actual accretion rate (orange dashed) following the minimum of the two. 
    (d) Eddington ratio $\lambdaEdd = \dot{M}/\dot{M}_{\rm Edd}$ evolution, showing periods of near-Eddington accretion at high redshift, followed by sub-Eddington growth at intermediate redshifts, then rise again at $z\sim 10$. 
    Blue and green shaded regions in all panels represent $1\sigma$ uncertainties derived from the Monte Carlo parameter inference. The probability of black hole seed formation is extremely low at $z > 30$, indicated by the left side of the yellow vertical line. 
    }
    \label{fig:accretion_fit_J0313_1806}
\end{figure*}

\textbf{J0313-1806} is a luminous quasar discovered at $z=7.642$ \citep{2021ApJ...907L...1W, 2021ApJ...923..262Y}. Initial spectroscopic observations with Magellan/FIRE confirmed the high-redshift nature by revealing a clear Lyman break at 1.048 $\mu$m, indicating a redshift of $z \gtrsim 7.6$ \citep{2010SPIE.7735E..14S}. Further ALMA observations provided a precise systemic redshift of $z_{\rm [CII]}=7.6423 \pm 0.0013$ based on the detection and spectral fitting of the [C II] emission line \citep{2021ApJ...907L...1W}. The quasar's bolometric luminosity was determined through spectral fitting of the combined and flux-calibrated near-infrared spectra. A pseudo-continuum model, comprising a power-law, Fe II emission, and Balmer continuum, was fit to spectral regions free of strong lines. By applying a bolometric correction factor of $C_{3000}=5.15$ \citep{2011ApJS..194...45S} to the continuum luminosity measured at rest-frame 3000 \AA, the bolometric luminosity was calculated to be $(1.4 \pm 0.1) \times 10^{47}$ erg s$^{-1}$ (or $3.6 \times 10^{13} L_\odot$) \citep{2021ApJ...907L...1W}. The mass of the central supermassive black hole (SMBH) in J0313-1806 was estimated using the widely utilized Mg II virial estimator \citep{2009ApJ...699..800V}. This method involves analyzing the width of the broad Mg II emission line, which is related to the velocity dispersion of the gas in the broad-line region gravitationally bound to the black hole. After subtracting the fitted pseudo-continuum model, a two-Gaussian model was fit to the extracted Mg II line profile. This fitting procedure yielded a full width at half maximum (FWHM) of $3670 \pm 405$ km s$^{-1}$ for the Mg II line, leading to an estimated SMBH mass of $(1.6 \pm 0.4) \times 10^{9} M_\odot$ \citep{2021ApJ...907L...1W}.

\begin{figure*}
    \centering
    \includegraphics[width=1\linewidth]{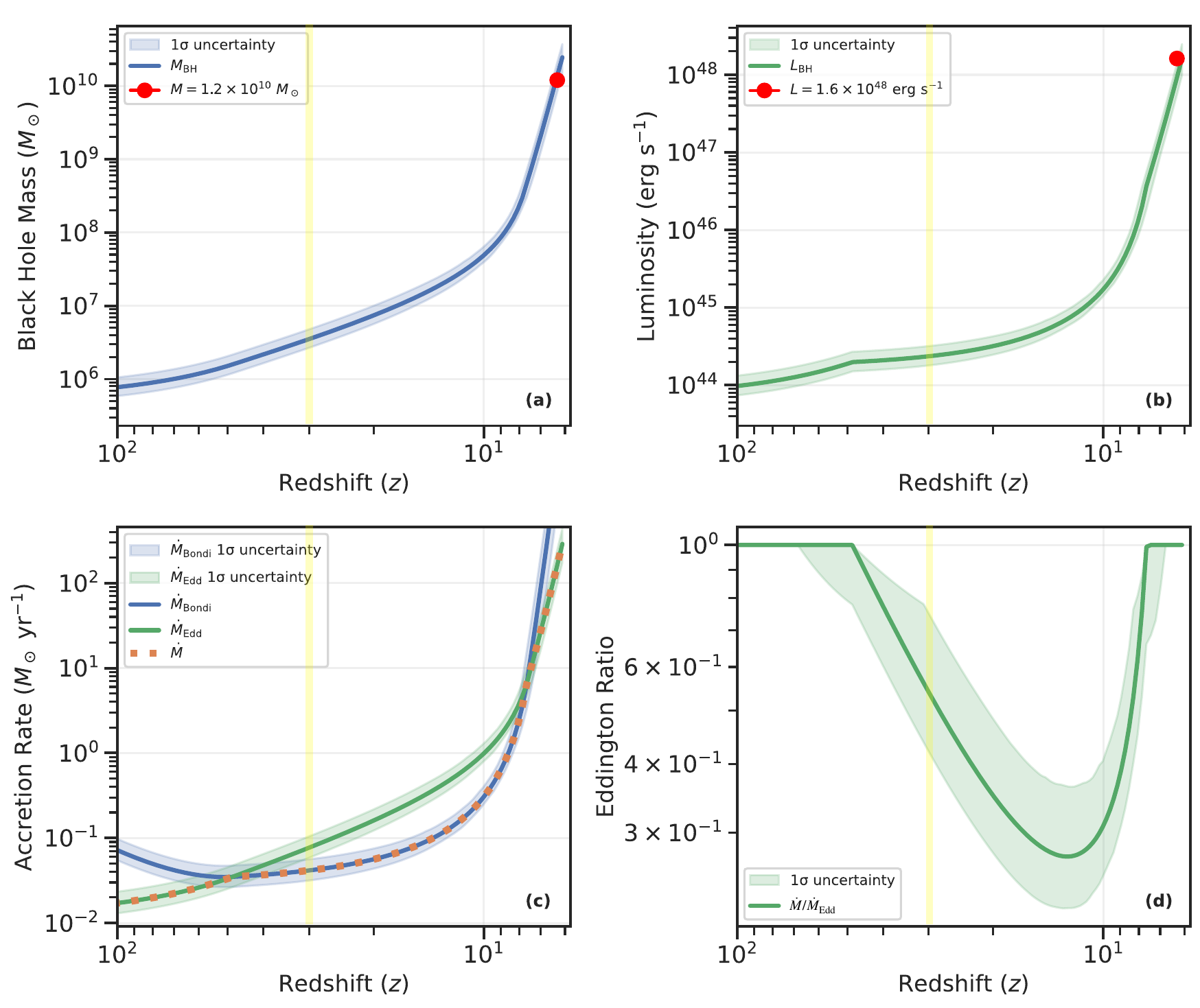}
    \caption{
\textbf{Evolutionary history of the quasar J010013.02+280225.8.} 
(a) Black hole mass growth, starting from $8.0^{+2.6}_{-2.1} \times 10^5\, M_\odot$ at $z = 100$ to $1.2 \times 10^{10}\, M_\odot$ at $z = 6.30$ in a dense environment with $\rho_0 / \rho_{\mathrm{b}} = 78.88^{+32.21}_{-23.49}$. 
(b) Corresponding luminosity evolution, reaching $1.6 \times 10^{48}$ erg s$^{-1}$. 
(c) Accretion rate history showing the transition between Bondi-limited (blue) and Eddington-limited (green) regimes, with the actual accretion rate (orange dashed) following the minimum of the two. 
(d) Eddington ratio $\lambdaEdd = \dot{M}/\dot{M}_{\rm Edd}$ evolution, showing periods of near-Eddington accretion at high redshift, followed by sub-Eddington growth at intermediate redshifts, then rise again at $z\sim 10$. 
Blue and green shaded regions in all panels represent $1\sigma$ uncertainties derived from the Monte Carlo parameter inference. The probability of black hole seed formation is extremely low at $z > 30$, indicated by the left side of the yellow vertical line. 
}
\label{fig:accretion_fit_J010013_02_280225_8}
\end{figure*}

\vspace{1em}

\textbf{SDSS J010013.02+280225.8} (J0100+2802) is presented as an ultraluminous quasar discovered at a redshift of $z=6.30 \pm 0.01$ \citep{2015Natur.518..512W}. Its initial identification came from optical and near-infrared color selection using data from SDSS, 2MASS, and WISE surveys \citep{2000AJ....120.1579Y,2006AJ....131.1163S,2010AJ....140.1868W}. The redshift was spectroscopically confirmed through observations with the Lijiang 2.4-m, MMT, and LBT telescopes, which revealed a sharp break in the spectrum and the prominent Lyman-alpha emission line shifted to approximately $8900$ \AA\ \citep{2015Natur.518..512W,2024ApJ...968..118P}. Its optical luminosity at rest-frame 3000 \AA\ was measured from spectroscopy to be $(3.156 \pm 0.47) \times 10^{47}$ erg s$^{-1}$. This optical luminosity was then used to estimate the bolometric luminosity by applying an empirical conversion factor, resulting in $L_{\rm bol} = 1.62 \times 10^{48}$ erg s$^{-1}$ \citep{2015Natur.518..512W}. The black hole mass was estimated using the virial method based on near-infrared spectroscopy (from Gemini and Magellan telescopes) of the broad Mg II emission line \citep{1999ApJ...526..579W,2004MNRAS.352.1390M}. The full-width at half-maximum (FWHM) of the Mg II line was measured at $5130 \pm 150$ km s$^{-1}$. By combining this line width with the continuum luminosity at rest-frame 3000 \AA\, a mass of $(1.24 \pm 0.19) \times 10^{10} M_\odot$ was derived \citep{2015Natur.518..512W}.

\subsection{Fittings Methods}\label{subsec:fitting-methods}

To constrain the initial black hole seed mass $M_0$ and the density ratio factor $f_\rho$, we employ a Bayesian Markov Chain Monte Carlo (MCMC) approach \citep{gamerman2006markov}. We define the posterior probability as:

\begin{equation}
P(M_0, f_\rho | M_{\rm obs}, L_{\rm obs}) \propto P(M_{\rm obs}, L_{\rm obs} | M_0, f_\rho) \times P(M_0, f_\rho)
\end{equation}

The likelihood function assumes log-normal uncertainties in both mass and luminosity:

\begin{equation}
\begin{aligned}
\ln P(M_{\rm obs}, L_{\rm obs} | M_0, f_\rho) = -\frac{1}{2} \Bigg[ &\frac{(\log_{10}(M_{\rm model}/M_{\rm obs}))^2}{\sigma_M^2} \\
&+ \frac{(\log_{10}(L_{\rm model}/L_{\rm obs}))^2}{\sigma_L^2} \Bigg]
\end{aligned}
\end{equation}

where $M_{\rm model}$ and $L_{\rm model}$ are the model predictions for a given set of parameters, and $\sigma_M$ and $\sigma_L$ represent the assumed uncertainties in the logarithmic mass and luminosity measurements, respectively.

We adopt uniform priors within physically motivated ranges: $2 \leq \log_{10}(M_0/M_\odot) \leq 8$ and $0 \leq f_\rho \leq 250$. The MCMC sampling is performed using the emcee package \citep{2013PASP..125..306F} with 32 walkers and 5000 steps, discarding the first 20\% as burn-in. We verify convergence by examining the autocorrelation time of the chains.

For each set of parameters in our MCMC chains, we evolve the black hole from its initial seed and calculate the resulting mass and luminosity at the observed redshift. The posterior distributions of these parameters reveal the most probable formation pathway for this extreme object, while the spread in these distributions quantifies the uncertainty in our inferences.

By reconstructing the full evolutionary track of the black hole, including its mass growth, accretion rate, luminosity, and Eddington ratio as functions of redshift, we can identify phases in its history and determine whether its growth was primarily Bondi-limited or Eddington-limited throughout cosmic time.

\subsection{Results: Evolutionary Phases of Accretion}\label{subsec:results}

Table \ref{tab:smbh_properties_updated} shows the initial black hole mass required at different redshifts for each quasar to reach its observed mass and luminosity. According to the fermion model, the expected seed mass for black hole formation is in the order of $10^6 M_\odot$. By comparing these two cases, we find that a seed mass of approximately $10^6 M_\odot$ corresponds to redshift $z \approx 20-30$, while a mass equal to the lower limit inferred from the fermion model for Sgr A$^*$, $4.3 \times 10^6 M_\odot$, is reached at $z \sim 20$. This suggests that most early quasars grew from seeds with masses of order $10^6 M_\odot$ that formed around $z \sim 20$.

In the following discussion, as well as in the resulting Figures \ref{fig:accretion_fit_J0313_1806} and \ref{fig:accretion_fit_J010013_02_280225_8}, we adopt a black hole growth history starting from redshift $z = 100$. This choice allows for a complete description of the possible multiple episodes of growth that may occur in the early universe. However, it should be noted that the probability of black hole seeds forming at $z > 30$ is extremely low, and the majority of black holes will not experience all the following phases.

\begin{table*}[]
  \centering
  \caption{Properties of black hole seeds inferred when the growth calculation is initialized at different redshifts. The formation of massive seeds is expected to become appreciable only at $z\simeq20{-}30$, while the probability of producing such seeds at much higher redshift is small.} 
  \label{tab:smbh_properties_updated}
%  \resizebox{\textwidth}{!}{% Resize table to fit within text width
  \begin{tabular}{l cccccc}
    \toprule
\textbf{} & $\mathbf{z=1000}$ & $\mathbf{z=100}$ & $\mathbf{z=50}$ & $\mathbf{z=30}$ & $\mathbf{z=20}$ & $\mathbf{z=10}$ \\
    \toprule
    \multicolumn{7}{l}{\textbf{J0313-1806}} \\
    \midrule
    Mass (\Msun)       & $5.2^{+4.1}_{-2.0} \times 10^{5}$ & $7.5^{+5.9}_{-2.9} \times 10^{5}$ & $1.5^{+1.2}_{-0.6} \times 10^{6}$ & $3.4^{+2.7}_{-1.3} \times 10^{6}$ & $7.5^{+5.6}_{-2.9} \times 10^{6}$ & $6.4^{+2.8}_{-2.4} \times 10^{7}$ \\
    Luminosity (erg/s) & $6.5^{+5.2}_{-2.5} \times 10^{43}$ & $9.4^{+7.5}_{-3.6} \times 10^{43}$ & $1.8^{+1.4}_{-0.7} \times 10^{44}$ & $2.4^{+1.8}_{-0.9} \times 10^{44}$ & $3.4^{+2.4}_{-1.3} \times 10^{44}$ & $3.0^{+1.3}_{-1.1} \times 10^{45}$ \\
    Accretion Rate (\Msun/yr) & $1.2^{+0.9}_{-0.4} \times 10^{-2}$ & $1.7^{+1.3}_{-0.6} \times 10^{-2}$ & $3.2^{+2.5}_{-1.2} \times 10^{-2}$ & $4.2^{+3.1}_{-1.6} \times 10^{-2}$ & $5.9^{+4.1}_{-2.3} \times 10^{-2}$ & $5.3^{+2.4}_{-1.9} \times 10^{-1}$ \\
    Eddington Ratio ($\lambdaEdd$) & $1.0^{+0.0}_{-0.0}$ & $1.0^{+0.0}_{-0.0}$ & $1.0^{+0.0}_{-0.32}$ & $0.55^{+0.22}_{-0.18}$ & $0.37^{+0.32}_{-0.12}$ & $0.41^{+0.28}_{-0.11}$ \\
    \midrule
    \multicolumn{7}{l}{\textbf{J010013.02+280225.8}} \\
    \midrule
    Mass (\Msun)       & $5.2^{+2.4}_{-1.5} \times 10^{5}$ & $7.4^{+3.4}_{-2.2} \times 10^{5}$ & $1.4^{+0.7}_{-0.4} \times 10^{6}$ & $3.4^{+1.5}_{-1.0} \times 10^{6}$ & $6.9^{+3.1}_{-1.9} \times 10^{6}$ & $4.6^{+1.8}_{-1.3} \times 10^{7}$ \\
    Luminosity (erg/s) & $6.5^{+3.0}_{-1.9} \times 10^{43}$ & $9.3^{+4.2}_{-2.8} \times 10^{43}$ & $1.8^{+0.8}_{-0.5} \times 10^{44}$ & $2.2^{+1.0}_{-0.6} \times 10^{44}$ & $3.0^{+1.3}_{-0.8} \times 10^{44}$ & $1.6^{+0.6}_{-0.4} \times 10^{45}$ \\
    Accretion Rate (\Msun/yr) & $1.1^{+0.5}_{-0.3} \times 10^{-2}$ & $1.6^{+0.7}_{-0.5} \times 10^{-2}$ & $3.2^{+1.4}_{-0.9} \times 10^{-2}$ & $3.9^{+1.8}_{-1.1} \times 10^{-2}$ & $5.3^{+2.3}_{-1.4} \times 10^{-2}$ & $2.9^{+1.0}_{-0.7} \times 10^{-1}$ \\
    Eddington Ratio ($\lambdaEdd$) & $1.0^{+0.0}_{-0.0}$ & $1.0^{+0.0}_{-0.0}$ & $1.0^{+0.0}_{-0.25}$ & $0.54^{+0.23}_{-0.14}$ & $0.35^{+0.18}_{-0.09}$ & $0.31^{+0.13}_{-0.07}$ \\
    % \midrule
    % \multicolumn{7}{l}{\textbf{UHZ-1}} \\
    % \midrule
    % Mass (\Msun)       & $8.7^{+10.1}_{-5.2} \times 10^{5}$ & $1.3^{+1.4}_{-0.8} \times 10^{6}$ & $2.4^{+2.8}_{-1.5} \times 10^{6}$ & $5.7^{+6.5}_{-3.3} \times 10^{6}$ & $1.2^{+1.3}_{-0.7} \times 10^{7}$ & $1.1^{+0.6}_{-0.4} \times 10^{8}$ \\
    % Luminosity (erg/s) & $1.1^{+1.3}_{-0.7} \times 10^{44}$ & $1.6^{+1.8}_{-0.9} \times 10^{44}$ & $3.1^{+3.5}_{-1.8} \times 10^{44}$ & $3.8^{+4.2}_{-2.1} \times 10^{44}$ & $5.6^{+4.7}_{-3.1} \times 10^{44}$ & $4.3^{+7.1}_{-1.8} \times 10^{45}$ \\
    % Accretion Rate (\Msun/yr) & $1.9^{+2.2}_{-1.2} \times 10^{-2}$ & $2.8^{+3.2}_{-1.7} \times 10^{-2}$ & $5.4^{+6.2}_{-3.2} \times 10^{-2}$ & $6.7^{+7.4}_{-3.8} \times 10^{-2}$ & $9.9^{+8.2}_{-5.4} \times 10^{-2}$ & $7.6^{+12.4}_{-3.2} \times 10^{-1}$ \\
    % Eddington Ratio ($\lambdaEdd$) & $1.0^{+0.0}_{-0.0}$ & $1.0^{+0.0}_{-0.1}$ & $1.0^{+0.0}_{-0.40}$ & $0.55^{+0.23}_{-0.22}$ & $0.36^{+0.32}_{-0.14}$ & $0.41^{+0.30}_{-0.14}$ \\
    \bottomrule
  \end{tabular}
% } % End resizebox
\end{table*}

The growth trajectory of supermassive black holes from massive seeds ($M_0 \sim 10^6 M_\odot$), such as those potentially formed via fermion core collapse, is shaped by the evolving interplay between the black hole's mass accumulation, $M_{\text{BH}}(t)$, and the declining ambient baryon density, $\rho_b(t)$, driven by cosmic expansion. This dynamic competition dictates the dominant accretion mechanism and can be effectively analyzed by tracking the ratio $R = \dot{M}_B / \dot{M}_{\text{Edd}}$, which scales as $R \propto M_{\text{BH}}(t) \rho(t)$, relative to unity. Here, the Bondi accretion rate $\dot{M}_B$ is sensitive to local density and the black hole mass, the local density is parameterized as $\rho = f_\rho \rho_{b}(t)$, where $f_\rho \gg 1$ represents a sustained local overdensity crucial for efficient feeding. The Eddington limit $\dot{M}_{\text{Edd}}$ determines the Salpeter e-folding timescale $t_S = M_{\text{BH}}/\dot{M}_{\text{Edd}} \approx 45$ Myr for the efficiency $\eta=0.1$. 

\vspace{1em}

\textbf{Initial Eddington Phase:} In the earliest phase, at redshifts $z \gtrsim 40$ ($t \lesssim 80$ Myr), the universe is extremely dense $\rho_{b} \propto (1+z)^3$, ensuring that even with moderate $M_{\text{BH}}$, the product $M_{\text{BH}}\rho$ results in $R \gg 1$. Under these conditions, accretion is limited by radiation pressure, proceeding at the Eddington rate ($\dot{M} = \dot{M}_{\text{Edd}}$) with $\lambda_{\text{Edd}}=1$, leading to exponential mass growth, that $M_{\text{BH}}(t) \approx M_0 e^{(t-t_{\text{seed}})/t_S}$. To understand how the ratio $R$ evolves during this phase, we consider its time derivative. The density decay $\rho_{b} \propto (1+z)^3 \propto t^{-2}$ in the matter-dominated era (approximately valid for $z \gg 1$), then $\ln R = \text{const} + \ln M_{\text{BH}} + \ln \rho \approx \text{const} + (t/t_S) - 2 \ln t$. Differentiating gives $d(\ln R)/dt = 1/t_S - 2/t$. The ratio $R$ decreases when $d(\ln R)/dt < 0$, which implies $1/t_S < 2/t$. Physically, this condition means that the fractional rate of density decline ($2/t$) exceeds the constant fractional rate of mass growth ($1/t_S$), causing their product $R \propto M_{\text{BH}}\rho$ to decrease. As redshift decreases towards $z=40$ (where $t \approx 80$ Myr $\approx 2 t_S$), this condition $t < 2t_S$ is met, and thus the ratio $R$ naturally begins to decrease, setting the stage for the transition.

\vspace{1em}

\textbf{Transitional Bondi Phase:} The critical transition occurs around $z \approx 40$ ($t \approx 2 t_S$) as specified by our model fitting. The accretion regime physically switches from being radiation-pressure limited to being supply-limited, governed by the Bondi formalism. The actual accretion rate becomes $\dot{M} = \dot{M}_B$, and the Eddington ratio $\lambda_{\text{Edd}}$ falls below 1. Following the transition, from $z \approx 40$ down towards $z \approx 15$, the ratio $R$  continues to decrease  before reaching its minimum value ($R \sim 0.3 - 0.5$) around $z \approx 15$. This persistence of the decreasing trend occurs because the fractional mass growth rate has changed from the constant $1/t_S$ to the generally smaller rate $\dot{M}_B/M_{BH} \propto R(t)$. This initially small growth rate allows the density decay term ($2/t$) to continue dominating the balance in $d(\ln R)/dt = (\dot{M}_B/M_{BH}) - 2/t$ for some time after the transition. Growth during this phase relies heavily on maintaining a high local density enhancement $f_\rho$ to counteract the cosmic dilution ($\rho_{b} \propto t^{-2}$). 

\vspace{1em}

\textbf{Later Eddington Phase:} Finally, at lower redshifts ($z \lesssim 15$), the accretion physics is now fundamentally different, governed by the Bondi rate $\dot{M}_B \propto M_{\text{BH}}^2 \rho$. In this regime, the absolute magnitude of the black hole mass $M_{\text{BH}}$, now enormous ($>10^8 M_\odot$), becomes the leading factor. The $M_{\text{BH}}^2$ dependence in the accretion rate means that further mass growth strongly amplifies the rate itself. This powerful effect starts to overwhelm the continuing, but relatively slower, decline in $\rho(t)$. Consequently, the product $M_{\text{BH}}(t) \rho(t)$, and thus the ratio $R = \dot{M}_B / \dot{M}_{\text{Edd}}$, reverses its trend and increases again. This resurgence is driven by the black hole's immense gravity (represented by $M_{\text{BH}}$) finally winning the tug of war against cosmic dilution ($\rho(t)$ decay) at late times. This causes $\lambda_{\text{Edd}}=R$ to rise back towards unity, powering the luminous quasar phase and enabling the final growth to the observed $10^9 - 10^{10} M_\odot$ scale, provided a robust gas supply ($f_\rho \gg 1$) is maintained.

\vspace{1em}

\textbf{Expected Long-Term Trend:} Although the model shows a rise in $R$ after $z \approx 15$ relevant for high-z quasars, the long-term evolution into the low-redshift universe ($z \ll 1$) would see $\dot{M}_B$ drops due to fuel exhaustion, which is not modeled in this article. This would eventually leads to a likely continues decrease in $R = \lambda_{\text{Edd}}$ towards the low values observed in quiescent SMBHs today.

\section{Comparison and Discussion}\label{sec:discussion}

In this section, we first compare with the traditional models of SMBH seeds.

If the first generation of black hole ``seeds'' formed as remnants of Population~III stars (with masses $M_{\rm seed} \sim 10^2$--$10^3\ M_\odot$), they would need to gain orders of magnitude in mass within a few hundred million years to reach $10^8$--$10^9\ M_\odot$ by $z > 7$. Even under optimal conditions of Eddington-limited accretion, the characteristic Salpeter time for doubling a black hole's mass is on the order of $\sim 4 \times 10^7$ years (assuming $\sim 10\%$ radiative efficiency). Growing from $10^2\ M_\odot$ to $10^9\ M_\odot$ requires a continuous Eddington accretion period of $\sim 0.7$ Gyr. This is a substantial fraction of the age of the Universe at $z \approx 7$ ($\sim 0.8$ Gyr), leaving very little margin for interruptions.  In practice, accretion may be sporadic or feedback-limited, so achieving such rapid growth is challenging. And in the extreme case of UHZ-1, even sustained Eddington accretion is insufficient to reach the observed black hole mass \citep{2024NatAs...8..126B}. One proposed solution is episodes of super-Eddington accretion, where the black hole accretes above the nominal Eddington limit for short periods. While super-Eddington growth can occur in gas-rich, dense environments, it is uncertain if it can be sustained long enough to bridge the gap.

Another challenge is forming sufficiently massive seeds so that less overall growth is needed. Besides stellar-mass remnants, theoretical models have proposed DCBHs as an alternate path \citep{1994ApJ...432...52L}. In a DCBH scenario, the gas cloud in a protogalactic halo collapses directly into a black hole of mass $M_{\rm seed} \sim 10^5$--$10^6 M_\odot$, skipping the normal star-formation stage. Achieving a direct collapse requires the gas to avoid fragmentation and cooling via $H_2$; this is possible if the halo's gas remains atomic (temperature $T \sim 10^4$ K) by either having very low metal enrichment and a strong Lyman-Werner ultraviolet background to dissociate $H_2$, or by dynamically inflowing so rapidly that star formation is suppressed \citep{2003ApJ...596...34B,2006MNRAS.371.1813L}. 

For the fermion dark matter, whether a self-gravitating fermion halo can truly form a super-Chandrasekhar core remains uncertain, but several physically consistent pathways have been proposed. Numerical work in the RAR model demonstrates that a semi-degenerate halo can evolve adiabatically, raising its central phase-space density while the outer envelope expands. Once the enclosed mass exceeds the critical value set by relativistic degeneracy pressure, the core is expected to collapse~\citep{2024arXiv241217919C}. Gravothermal contraction due to self-interacting dark matter may provide an alternative route by concentrating mass within a cosmologically relevant timescale, much shorter than two-body relaxation in a purely collisionless system~\citep{2021MNRAS.500.2177C}. Another possibility is that enhanced small-scale power or an early matter-dominated phase produces rare overdensities that already surpass the Chandrasekhar threshold before conventional structure formation occurs \citep{2012PhRvD..85l5027B, 2021PhRvD.103j3514B}.

An important advantage of the fermion black hole seed model is that it provides a well-defined physical scale for the initial black hole mass, set by the degeneracy pressure of the fermionic dark matter. This natural mass threshold, around $10^6 M_\odot$, removes the need for fine-tuning or extreme initial conditions required in other scenarios. A further advantage of starting from a massive seed is that the timescale to grow to the observed supermassive black hole mass is significantly reduced. This alleviates the strong time constraints that challenge models beginning with much lower mass seeds. With a $10^6 M_\odot$ initial mass, the black hole can reach $10^9$–$10^{10} M_\odot$ by $z > 6$ through more natural accretion modes.

We also estimate the detectability of high redshift seeds by JWST.  A quasar with luminosity $L=3\times10^{44}\,\mathrm{erg\,s^{-1}}$ at $z=20$ has a bolometric flux of $4.7\times10^{-17}\,\mathrm{erg\,s^{-1}\,cm^{-2}}$, while one with $L=2\times10^{44}\,\mathrm{erg\,s^{-1}}$ at $z=30$ has $1.3\times10^{-17}\,\mathrm{erg\,s^{-1}\,cm^{-2}}$.  After applying a UV bolometric correction and accounting for wide filter bandwidths and absorption, these correspond to AB magnitudes $\sim29$ at $z=20$ and $\gtrsim30.5$ at $z=30$.  Detection at $z\sim20$ would require ultra deep exposures, and at $z\sim30$ it would likely be feasible only with strong gravitational lensing or extremely long integration times. Consequently, standard JWST observing programs are unlikely to detect seed accretion directly at these redshifts.

\section{Conclusions}\label{sec:conclusion}

We have reconstructed the growth histories of two well-measured high-redshift quasars using a minimal accretion model in which the inflow rate is set by $\dot{M}=\min(\dot{M}_B,\dot{M}_{\mathrm{Edd}})$. The fits favor massive initial seeds, $M_0 \sim 10^{6}\,M_\odot$, and dense environments with baryonic overdensity factors $f_\rho \gtrsim 50$, with the most plausible seed-formation window at $z\sim20{-}30$. Under these conditions, the black hole can reach $>10^{9}\,M_\odot$ by $z\sim7$ without requiring continuous Eddington growth or any super-Eddington phase. In our solution, the accretion history is supply-limited for most of the available time and becomes near-Eddington only when the combination of black hole mass and gas density is large enough.

The main observational motivation for adopting this simple, cosmology-based framework comes from the newly observed LRDs, which suggest that black holes with masses of order $10^{6}\,M_\odot$ can already be active when the surrounding system is still compact and dominated by primordial gas, and before a mature galaxy structure, such as a settled disk and bulge, is in place. This motivates revisiting early AGN growth with a minimal set of inputs that track the evolving gas supply set by the expanding Universe. In this sense, the purpose of the present work is to quantify whether pre-galactic seeds in overdense regions, as implied by LRD phenomenology, can naturally account for the observed high-redshift quasar masses.

Two follow-up studies will extend the present work. The first will use cosmological perturbation theory to compute when and where fermion core collapse produces massive seeds, and to show that the formation probability becomes significant around $z\sim20{-}30$, consistent with the redshift range preferred by our fits. The second will develop a unified picture in which both AGN and LRD originate from the same seed-plus-accretion mechanism, with their different observed properties primarily reflecting different initial masses and local overdensities, therefore different supply-limited growth histories.

%The scenario is self-consistent: quantum-degenerate fermion dark matter provides a physical mass scale for the seed, and subsequent accretion always follows the minimum of physically motivated limits as the ambient density decreases with cosmic expansion. The same equilibrium model of fermions that describes both flat galaxy rotation curves and the S-star trajectories also allows for rapid early black hole growth, supporting a unified framework for galaxy dark matter structure and the emergence of the first supermassive black holes.

\section*{Acknowledgments}
We thank Professors C. R. Argüelles, J. A. Rueda, and G. V. Vereshchagin for helpful discussion and comments.

\bibliographystyle{cas-model2-names}
%\bibliographystyle{aasjournal}

%\bibliography{AGN-LRD-resub1}

\end{document}